\documentclass[aps, prb, preprint, superscriptaddress, a4paper, 11pt, floatfix,usenames,dvipsnames]{revtex4-1}
\usepackage[utf8]{inputenc}
\usepackage{amsmath}
\usepackage[colorlinks=true, linkcolor=black,citecolor = black,filecolor = black]{hyperref}
\usepackage{cleveref}

\usepackage{float}
\usepackage{graphicx}
\usepackage{epstopdf}
\usepackage[multi-part-units = single]{siunitx}
\usepackage{lineno}
\usepackage[export]{adjustbox}
\usepackage{comment}
\usepackage{xcolor}
\usepackage{nicefrac}

\usepackage{soul}
\usepackage{silence}
\usepackage{colortbl}

\usepackage[a4paper,margin=1in]{geometry}

\begin{document}
\raggedbottom
\title{Attosecond correlation interferometry}

\author{
    Assaf~Shonfeld$^{1,5\dagger}$,
    Keren~Deutsch$^{1\dagger}$,
    Noa~Yaffe$^{1}$,
    Michael~Birk$^{2}$,
    Matan~Even~Tzur$^{3}$,
    Barak Dayan$^{5}$,
    Oren~Cohen$^{2,4}$,
    Nirit~Dudovich$^{1*}$,
    Chen~Mor$^{1*}$ \\
	\small$^{1}$Department of Physics of Complex Systems, Weizmann Institute of Science, Rehovot, Israel\\
    \small$^{2}$ Technion - Israel Institute of Technology, Haifa, Israel\\
    \small$^{3}$Max Planck Institute for the Structure and Dynamics of Matter, Hamburg, Germany\\
    \small$^{4}$  Guangdong Technion-Israel Institute of Technology, Shantou, Guangdong 515063, China\\
    \small$^{5}$Department of Chemical and Biological Physics, Weizmann Institute of Science, Rehovot, Israel\\
	\small$^\ast$Corresponding authors. Email: chen.mor@weizmann.ac.il\\
    \small Email: nirit.dudovich@weizmann.ac.il\\
	\small$^\dagger$These authors contributed equally to this work.
}

\begin{abstract}

Correlations between optical modes, ranging from classical fluctuations to quantum entanglement, are a cornerstone of modern optics and emerging quantum technologies. However, probing these correlations on the natural timescale of electronic motion -- the attosecond regime -- remains a major challenge. Here, we generate and characterize correlated attosecond extreme-ultraviolet (XUV) emission by perturbing gas-phase high-harmonic generation with two-mode bright squeezed vacuum. The signal and idler fields imprint their correlated amplitude and phase fluctuations onto two families of harmonics, thereby transferring these correlations from the infrared to broadband XUV modes. Even harmonics serve as an intrinsic attosecond phase-sensitive probe of these correlations through interference between indistinguishable XUV pathways. Shot-resolved covariance measurements reveal the amplitude correlations, while even-harmonics resolve the phase correlations with sub-cycle precision. Time-domain reconstruction of the correlated emission reveals a transition from maximal fluctuations to near-silence within only 30 as. Whereas the signal and idler driven attosecond pulse trains exhibit excess fluctuations individually, their joint reconstruction shows strongly suppressed fluctuations, reflecting their shared correlations. These results establish a framework for generating and probing quantum correlations on attosecond timescales, opening the door to quantum attosecond science.

\end{abstract}

\maketitle

Correlations play a central role in modern physics, from classical statistical ensembles to the foundations of quantum mechanics. Since the early studies of quantum correlations, non-classical phenomena involving entangled quantum states have been uncovered, ranging from Bell-inequality violations and quantum non-locality\cite{bell1964einstein} to emergent solid-state phenomena\cite{amico2008entanglement}. Beyond their foundational role, quantum correlations have become a central resource for information processing, enabling advances in quantum communication\cite{yin2020entanglement}, computation\cite{arute2019quantum} and precision metrology\cite{nagata2007beating}. In the classical limit, correlations provide a powerful framework for noise suppression\cite{frasinski2016covariance}, enhanced measurement sensitivity\cite{nanograv2023evidence} and improved resolution\cite{katz2014non}. Approaches such as Hanbury Brown–Twiss interferometry\cite{brown1956correlation} and ghost imaging\cite{erkmen2010ghost} exploit statistical dependencies to access information beyond mean intensity or phase. Pioneering experiments have extended correlation measurements to the X-ray regime \cite{singer2013hanbury,pelliccia2016experimental,gorobtsov2018seeded,driver2020attosecond,Krebs2025XPDC}. In the temporal domain, correlations provide direct access to dynamical evolution, linking time-dependent fluctuations to the underlying interactions and degrees of freedom. Advances in ultrafast science have enabled measurements of quantum correlations between electrons and ions\cite{blaga2012imaging,wolter2016ultrafast} and optical correlations on femtosecond timescales\cite{mukamel2000multidimensional}, revealing vibrational motion\cite{hamm1998structure}, energy transfer and decoherence in free electrons\cite{Meier2023FewElectron}, molecules\cite{engel2007evidence} and solids\cite{weiss2023discovery}. Extending optical correlation measurements to the natural timescale of electronic motion—the attosecond regime—remains an outstanding challenge. While attosecond science has achieved remarkable progress in resolving sub-cycle electron dynamics\cite{krausz2009attosecond}, optical correlations within attosecond emission have remained elusive.

Accessing quantum correlations on attosecond timescales holds the potential to reveal the quantum nature of ultrafast light and matter dynamics. Indeed, attosecond photoionization experiments have resolved and controlled such correlations in matter~\cite{koll2022experimental,laurell2025measuring,koll2026entanglement}. The next conceptual leap is to bring attosecond correlations into the optical domain.  At the heart of attosecond science is high-harmonic generation (HHG), a highly nonlinear light–matter interaction that generates attosecond pulses\cite{Li1989,paul2001observation}. Pioneering studies of HHG have demonstrated non-classical correlations between the IR driving field and the high harmonics \cite{tsatrafyllis_high-order_2017,Lewenstein2021} and between harmonics across the visible spectral range on femtosecond timescales\cite{theidel2024evidence,theidel2025observation}. Recently, amplitude correlations were generated between UV high harmonics via an entangled photons source\cite{lyu2026attosecond}. Together with theoretical predictions of multiple mechanisms capable of generating quantum-correlated harmonics\cite{Sloan2023,Yi2025,Lange2024,rivera2025structured}, these findings establish HHG as a promising platform for exploring correlations in strongly nonlinear regimes and on sub-cycle timescales.  Can such correlations be extended into the extreme ultraviolet (XUV)? Can their structure be controlled with attosecond precision? Bringing the fundamental concept of correlated photons into the attosecond regime poses two key challenges: generating correlated photons within attosecond bursts and detecting their correlations with sub-cycle temporal resolution.

Here, we generate correlated attosecond optical pulses and resolve their amplitude and phase correlations with sub-cycle precision. We achieve this by combining IR twin beams -- signal and idler modes generated in a two-mode bright squeezed vacuum (TMBSV) state\cite{chekhova2015bright} -- with a strong classical driving field to induce HHG, enabling the transfer of amplitude and phase correlations into the XUV regime. This process gives rise to fractional harmonics, which map the correlations of the IR modes onto broadband correlations between distinct XUV spectral components. A central element of our approach is the ability to resolve these correlations with attosecond precision. The appearance of even harmonics acts as an intrinsic attosecond coincidence interferometer, providing access to temporal correlations between distinct XUV fractional modes. By scanning the delay between the TMBSV and the driving field, we observe pronounced oscillations in the even-harmonic intensity, reflecting phase and amplitude correlations between the fractional harmonics. Tracing the fractional harmonics into the time domain shows that strong shot-to-shot fluctuations obscure the attosecond dynamics, whereas their joint temporal reconstruction reveals sub-cycle noise dynamics, with $\sim30$ attoseconds separating maximal noise from field silence.

A fundamental mechanism for generating quantum correlations in the visible and IR regimes is non-degenerate spontaneous parametric down-conversion (SPDC). In the frequency non-degenerate case, the signal and idler modes are spectrally distinct and obey energy conservation, $\omega_s+\omega_i=\omega_p$, where $\omega_p$ is the fundamental frequency and $\omega_s$, $\omega_i$ are the central frequencies of the signal and idler, respectively. At low photon flux, the process exhibits measurable entanglement between them\cite{ou1988violation}. At high gain, the process generates TMBSV consisting of macroscopic signal and idler fields with complex thermal fluctuations \cite{agafonov2010two, Iskhakov_Superbunched_2012}. Degenerate bright squeezed vacuum (BSV) has recently been employed in HHG, demonstrating the sensitivity of harmonic generation to the quantum statistics of the driving field~\cite{Rasputnyi2024,VampaSolidslemieux2025photon,tzur2025attosecond}. The frequency-non-degenerate configuration offers a key advantage: it maps the correlations between two spectrally distinct components onto separate harmonic channels. The pairwise parametric interaction imposes strong correlations in both photon number and phase:  $n_s^{(m)} = n_i^{(m)} \Rightarrow |E_s|^{(m)} = |E_i|^{(m)}$,
 $\phi_s^{(m)}+\phi_i^{(m)} \approx \phi_p$, where $n_{s,i}^{(m)}$ and $\phi_{s,i}^{(m)}$ denote the photon number and optical phase of the signal and idler modes for the $m$-th realization, respectively, while $\phi_p$ is the pump phase. These relations reflect two key constraints: correlated amplitude fluctuations arising from pairwise generation, and a fixed phase sum imposed by the parametric interaction\cite{tanas1996quantum, Iskhakov_Superbunched_2012, chekhova2015bright}. Both constraints persist at arbitrary parametric
gain, while the individual phases remain uniformly random (see Supplementary Information).

How can we transfer the concept of correlated photonic emission  into the attosecond domain? Attosecond pulses arise from the three-step mechanism \cite{corkum2007attosecond}. In this picture, an intense driving field (with frequency $\omega_p$) liberates an electron via tunnel ionization, accelerates it in the continuum, and drives it back to recombine with its parent ion, emitting an XUV burst with sub-cycle duration. Adding  a weak TMBSV field perturbs the HHG process by modifying both the tunneling step -- enhancing or suppressing the barrier -- and the electron trajectories, thereby altering the accumulated phase of the electronic wavefunction. Overall, the perturbation is imprinted as a complex phase: the imaginary component reflects changes in tunneling, while the real component arises from trajectory modifications\cite{salieres2001feynman,ivanov2005anatomy,Pedatzur2015,tzur2025attosecond}.
For a weak perturbation, each of the signal and idler fields contributes a complex phase to the emitted XUV field: $A^{XUV} \approx A^{XUV}_0\cdot e^{\sigma_{s}+\sigma_{i}}$. Here $A^{XUV}_0$ denotes the unperturbed XUV field, while $\sigma_s$ and $\sigma_i$ represent complex phase perturbations, proportional to the perturbation fields $E_s,$ $E_i$, respectively (see Supplementary Information).

Moving beyond the single-cycle picture reveals a richer temporal structure of the perturbation, which is time-periodic with periodicities set by $\omega_p$, $\omega_s$, and $\omega_i$. This temporal structure is mapped onto the HHG spectrum, leading to the appearance of new frequency components. When HHG is driven by a single-color field, symmetry between successive half-cycles results in the generation of odd harmonics only. Adding the correlated fields breaks this symmetry, giving rise to a longer-timescale structure dictated by the temporal beating between $\omega_p$, $\omega_s$, and $\omega_i$. In the frequency domain, these perturbations lead to the appearance of sidebands at fractional harmonic numbers $2N \pm \delta_s\,,\;2N \pm \delta_i\,;\; \delta_{s,i} \equiv \frac{\omega_{s,i}}{\omega_p}$, associated with the signal and idler fields, respectively.

To leading order, the fractional harmonic fields scale linearly with the perturbation fields:
$E_{2N + \delta_s}\propto E_s^*\,,\;E_{2N - \delta_s}\propto E_s$, and $E_{2N + \delta_i}\propto E_i^*\,,\;E_{2N - \delta_i}\propto E_i$
(see Supplementary Information). Hence, the fractional harmonics inherit the amplitude and phase fluctuations, as well as the correlations, of the signal and idler IR fields, mapping their intricate stochastic structure from the IR to the XUV.
\begin{equation}
\label{eq:frac harmonics}
n_{2N\pm\delta_s}^{(m)}
\sim
n_{2N\pm\delta_i}^{(m)},
\qquad
\phi_{2N\pm\delta_s}^{(m)}
+
\phi_{2N\pm\delta_i}^{(m)}
\approx
\mp\phi_p+\phi_{0}.
\end{equation}
where
$n_{2N\pm\delta_{s,i}}^{(m)}$ and
$\phi_{2N\pm\delta_{s,i}}^{(m)}$ denote the photon numbers and optical phases of the signal and idler driven fractional harmonics in the $m$-th stochastic realization, respectively. The phase sum follows the driving-field phase $\phi_p$, which is taken as the reference phase for all shots and therefore treated as constant, up to a realization-independent phase offset $\phi_{0}$ introduced by the HHG process.
 The theoretically squeezed quantum phase fluctuations, as well as residual technical noise are neglected, as they are small compared with the uniformly distributed shot-to-shot phase fluctuations considered here.

\begin{figure}[h!]
\centering
    \includegraphics[width=1\linewidth]{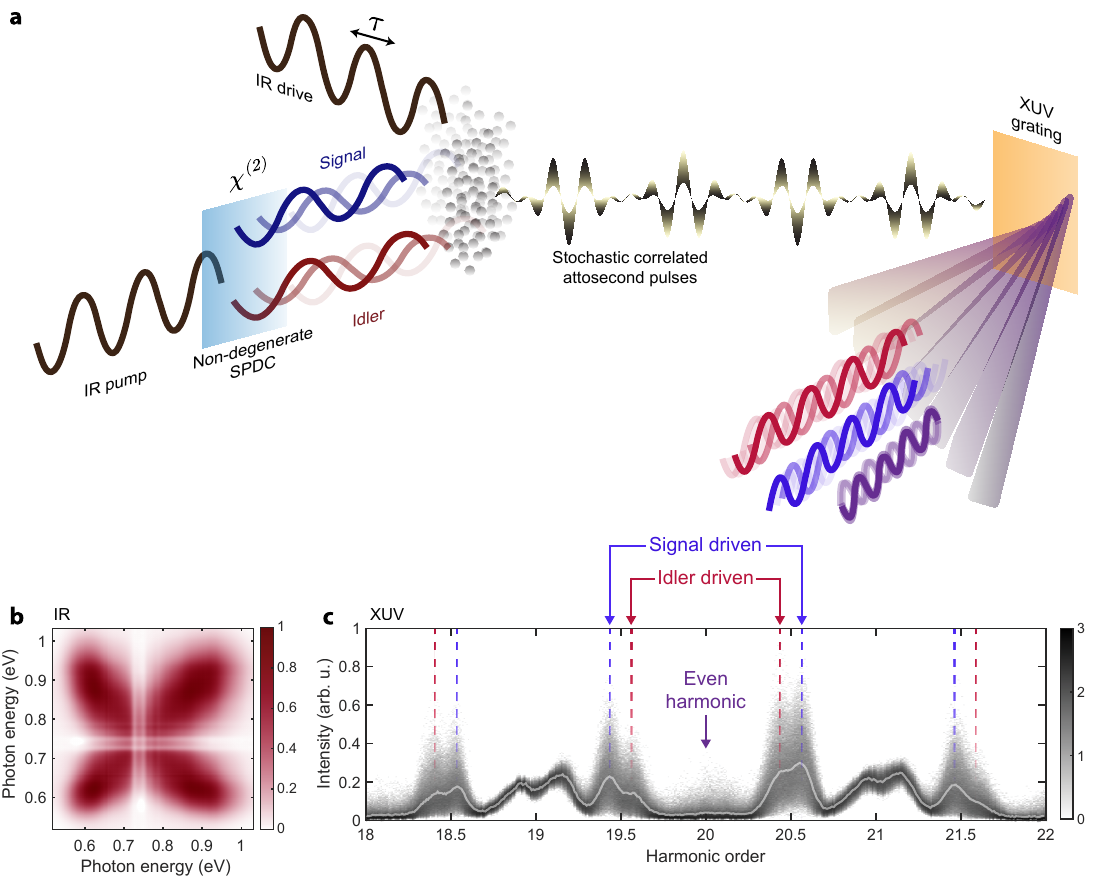}
    \caption{
    \textbf{Generation of Correlated Attosecond Pulses}
\textbf{(a)} Schematic of the experimental  setup.
An 800 nm laser pump  is divided into a strong coherent field that drives HHG and a pump field for non-degenerate SPDC, where signal (dark blue) and idler (dark red) modes of two-mode bright squeezed vacuum (TMBSV) state, are generated  in $\chi^{(2)}$ crystals. Together, they co-linearly perturb a coherent-field-driven HHG process, generating stochastic yet correlated attosecond pulse trains. A diffraction grating separates the XUV emission into signal driven (blue) and idler driven (red) fractional harmonics, which inherit the stochastic amplitude and phase fluctuations of the corresponding IR modes, and even harmonics, whose phase is locked to the driving field.
 \textbf{(b)}  Amplitude covariance matrix of the TMBSV perturbation field in the IR. The counter-diagonal four-lobed structure reveals the correlated amplitude fluctuations between the non-degenerate signal and idler modes. \textbf{(c)} Shot-to-shot globally normalized intensity histogram of the correlated XUV spectrum. Signal driven (blue) and idler driven (red) fractional harmonics, and even harmonic (purple) sideband groups emerge under TMBSV perturbation. The light gray line marks the mean intensity. The perturbative sidebands display pronounced long-tailed intensity statistics, characteristic of stochastic thermal fluctuations, whereas the odd harmonics retain the Gaussian intensity distribution of a coherent field.}  \label{fig:experimental_scheme}
\end{figure}

We reveal the emergence of correlated fluctuating attosecond light pulses by generating HHG in a three-field configuration, as illustrated in \Cref{fig:experimental_scheme}a. An IR pump field, centered at $\lambda_p = 800~\mathrm{nm}$ ($\omega_p = 1.55~\mathrm{eV}$), with a peak intensity of $\sim1\times10^{14}~\mathrm{W/cm^2}$, is used to generate TMBSV via SPDC, as well as to drive HHG in a krypton gas jet. The TMBSV consists of spectrally separated signal and idler modes, with central photon energies $\omega_i = 0.62~\mathrm{eV}$ and $\omega_s = 0.9~\mathrm{eV}$, corresponding to $\omega_i \approx 0.4\omega_p$ and $\omega_s \approx 0.58\omega_p$. The signal and idler frequencies satisfy $\omega_s+\omega_i=\omega_p^{\rm eff}$, where $\omega_p^{\rm eff}\approx1.52~\mathrm{eV}$, slightly red shifted from the central pump energy due to group velocity matching in the $\chi^{(2)}$ crystals
(see Supplementary Information).
The relative delay between the coherent pump and the TMBSV field is controlled to probe the correlation structure of the attosecond light. The XUV spectrum is recorded on a single-shot basis at each delay, resolving its fluctuations and correlations with attosecond precision. A detailed description of the experimental set-up is given in the Supplementary Information.

Single-shot measurements reveal the origin of the different harmonics in the HHG spectrum. The shot-to-shot XUV intensity histogram in \Cref{fig:experimental_scheme}c exhibits distinct spectral groups. The odd harmonics, driven by the coherent pump, display a narrow Gaussian intensity distribution. In contrast, the fractional harmonics exhibit long-tailed thermal intensity statistics, with a large standard deviation compared to their mean. We also observe even harmonics associated with higher-order perturbative processes, exhibiting increasingly large fluctuations, with standard deviations well exceeding the mean intensity. As will be discussed later, these harmonics play a crucial role in measuring the temporal correlations between harmonics, serving as an attosecond correlated interferometer.

\begin{figure}[h]
\centering
        \includegraphics[width=0.7\linewidth]{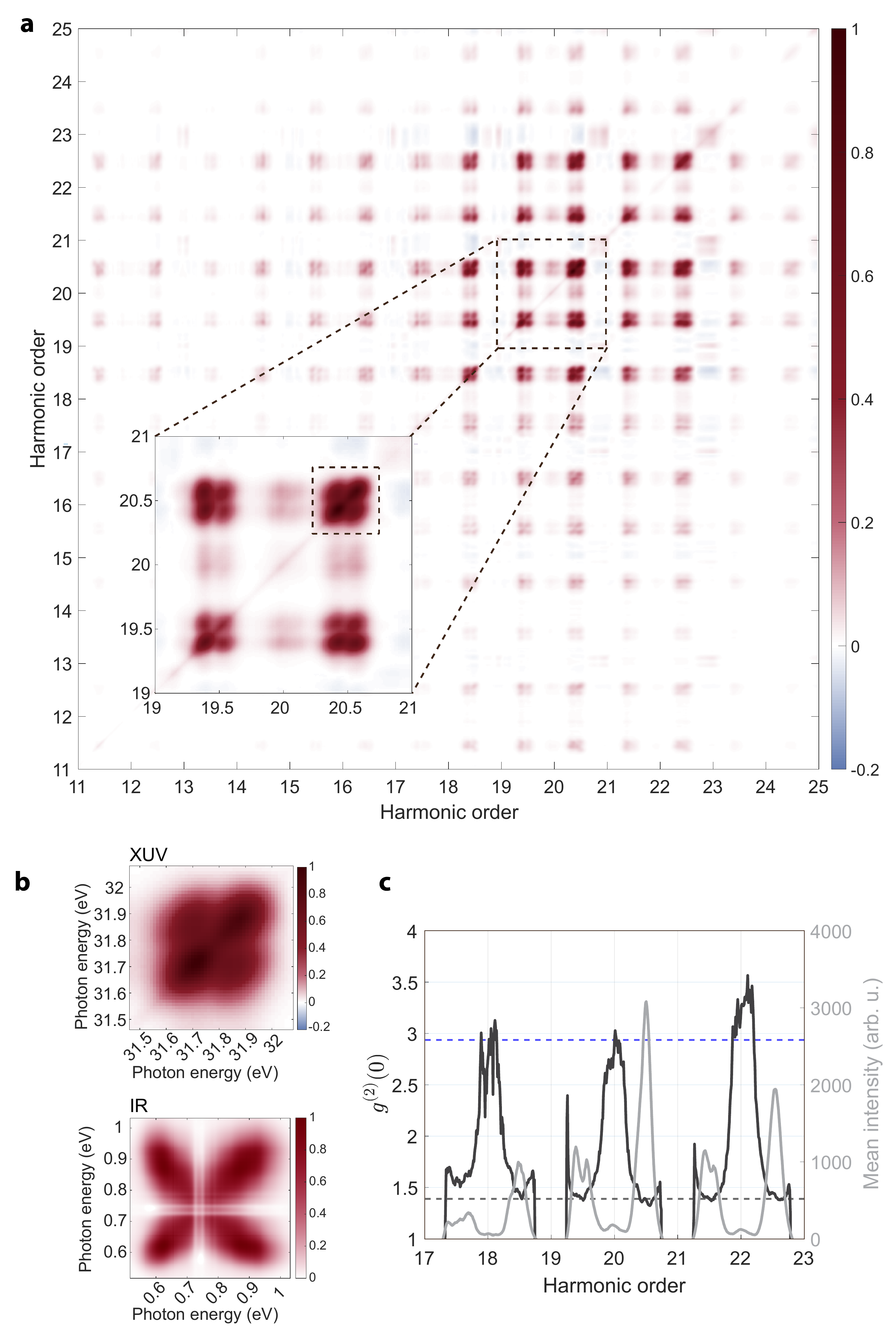}
    \caption{
\textbf{Broadband transfer of TMBSV amplitude correlations into XUV harmonics}
 \textbf{(a)} Shot-to-shot amplitude covariance matrix of the XUV spectrum generated under TMBSV perturbation, divided by the mean value squared and normalized to its maximum value for visibility. Correlations extend across the full harmonic bandwidth, linking signal and idler driven fractional harmonics. The inset highlights the covariance map of a single neighboring harmonic group. The four-lobed pattern reveals strong covariance between the signal and idler driven fractional harmonics (around harmonic orders 19.5 and 20.5), while the even harmonics exhibit covariance with both fractional-harmonic groups. \textbf{(b)} Upper panel, further enlargement of the boxed region in the inset of \textbf{(a)}. Lower panel, comparison with the IR TMBSV covariance matrix that reveals the same characteristic four-lobed covariance structure, demonstrating the transfer of the signal--idler amplitude correlations to the XUV. {\bf (c)} Mean intensity (gray) and $g^{(2)}(0)$ (black) of selected harmonics. The dashed gray (blue) line indicates the expected $g^{(2)}(0)$ value of the fractional (even) harmonics for fully correlated signal and idler driven fractional harmonics.}
\label{fig:amplitude_correlations}
\end{figure}

In the first stage, we study how spectral amplitude correlations are transferred from the IR TMBSV spectrum to the XUV HHG spectrum.
To this end, we analyze the spectral amplitude covariance, which quantifies the amplitude correlations between different frequencies in the spectrum: $\mathrm{Cov}\big(|E(\omega_1)|, |E(\omega_2)|\big) = \langle |E(\omega_1)| |E(\omega_2)| \rangle - \langle |E(\omega_1)| \rangle \langle |E(\omega_2)| \rangle.$ Figure~\ref{fig:experimental_scheme}b presents the spectral amplitude covariance matrix $\mathrm{Cov}\big(|E(\omega_i)|, |E(\omega_s)|\big)$, calculated from shot-to-shot  spectral measurements of the TMBSV field in the IR. The diagonal represents the variance of each frequency component, while the counter-diagonal structure reveals strong correlations between the non-degenerate signal and idler modes, forming a four-lobed pattern. Calculating the second-order coherence, $g^{(2)}(0) = 1 + \frac{\mathrm{Var}(I)}{\langle I \rangle^2}$, shows that each arm is individually thermal, with $g^{(2)}(0) \approx 2$ (See Supplementary Information).

We next demonstrate the transfer of the IR correlation structure to the XUV by the perturbed HHG process.  Figure~\ref{fig:amplitude_correlations}a presents the corresponding spectral amplitude covariance matrix, revealing a rich correlation structure between the sidebands. Neighboring fractional-harmonic pairs, $(2N\pm\delta_i, \quad 2N\pm \delta_s)$ (Figure~\ref{fig:amplitude_correlations}b), exhibit a characteristic four-lobed pattern that closely resembles the IR TMBSV correlations, indicating the transfer of the underlying correlation structure into the XUV. More importantly, the correlations are not confined to neighboring harmonics but extend across the entire $\sim 20$ eV bandwidth, linking all signal and idler driven sidebands. Their observation over such a broad bandwidth constitutes a key milestone of this work, providing direct evidence for the transfer of quantum-origin correlations from femtosecond to attosecond timescales.

Zooming into the covariance matrix (Figure~\ref{fig:amplitude_correlations}a, inset) reveals weak even harmonics that are not prominent in the mean spectrum. These features exhibit shot-to-shot covariance with both the signal and idler driven sideband groups, indicating that their generation arises from perturbative pathways involving contributions from both modes. The measured $g^{(2)}(0)$ values across the strongest XUV harmonics in the spectrum (Figure~\ref{fig:amplitude_correlations}c) provide further insight into the origin of the different harmonics. The signal and idler driven fractional harmonics exhibit $g^{(2)}(0)\approx1.4$, consistent with thermal fluctuations of the spectrally and spatially multi-mode beam in the experiment. In contrast, the even harmonics exhibit $g^{(2)}(0)\approx3.2$, reflecting the multiplicative nature of the underlying process, in which fluctuations of both modes combine to produce enhanced amplitude correlations (see Supplementary Information).

The harmonic emission comprises two spectrally distinct groups of fractional harmonics, $2N \pm \delta_s$ and $2N \pm \delta_i$, each forming an attosecond pulse train. Are these trains emitted simultaneously? What is their temporal interplay, and can their coincidence be resolved with attosecond precision? While we have so far established the transfer of spectral amplitude correlations from TMBSV to the HHG spectrum, addressing this question requires an additional conceptual step. It calls for the measurement of phase correlations between the emitted fields. This, in turn, requires an interferometric scheme with sub-cycle temporal resolution.

The generation of even harmonics arises from higher-order perturbative processes, in which both the signal and idler fields contribute within the same half-cycle of the driving field. While the fractional harmonics are dominated by a linear dependence on the TMBSV field, the even harmonics are governed by the joint contribution of the two modes, exhibiting a second-order dependence. To leading order, the even-harmonic field scales as $E_{\mathrm{2N}} \propto \sigma_s \sigma_i$ (see Supplementary Information). A unique property of these harmonics originates from their interferometric nature. The even-harmonic signal forms an interferometer comprising two indistinguishable pathways corresponding to two strong-field trajectories. For example, the generation of the $12^{\mathrm{th}}$ harmonic is governed by two quantum  pathways: absorption of a signal–idler photon pair from the $11^{\mathrm{th}}$ harmonic or emission of a pair from the $13^{\mathrm{th}}$ harmonic.  In general, the two path interference can be described as:

\begin{equation}
\label{eq:main:even harmonics two channels}
E^{(m)}_{2N} = E_{+}^{(m)}+E_{-}^{(m)}
\propto
\left|E_{2N\pm\delta_s}^{(m)}\right|\left|E_{2N\pm\delta_i}^{(m)}\right|
\left[
E_{2N+1}\,e^{\,i\left(\omega_p\tau+\Phi_{-}^{(m)}\right)}
+
E_{2N-1}\,e^{-i\left(\omega_p\tau-\Phi_{+}^{(m)}\right)}
\right]
\end{equation}
Here $\Phi_{\pm}^{(m)}\equiv\phi_{2N\pm\delta_s}^{(m)}
+\phi_{2N\pm\delta_i}^{(m)}$ are the single-shot phase sums of the
lower- and upper-sideband pairs of fractional harmonics,
\(E_{+}^{(m)}\) and \(E_{-}^{(m)}\) denote the two indistinguishable
pathways, \(E_{2N\pm1}\) are the complex amplitudes associated with the
neighboring odd harmonics, and \(\tau\) is the delay between the pump beam and the TMBSV beams, controlled with attosecond resolution.

The amplitudes $\left|E_{2N\pm \delta_s}^{(m)}\right| \propto \left|E_s^{(m)}\right|$ and $\left|E_{2N\pm \delta_i}^{(m)}\right|\propto \left|E_i^{(m)}\right|$ and the phases \(\phi_{2N\pm\delta_s}^{(m)}\) and \(\phi_{2N\pm\delta_i}^{(m)}\) correspond to the \(m\)-th realization of the signal and idler driven fractional harmonics fields respectively.

The mean value of the intensity of the even harmonics follows:
\begin{equation}
\begin{aligned}
\left\langle|E_{+}^{(m)}+E_{-}^{(m)}|^{2} \right \rangle
=
\left\langle DC \right\rangle
+
\left\langle 2\left|E_{+}^{(m)}\right|\left|E_{-}^{(m)}\right|
\cos\big(\Delta\Phi^{(m)} \big)\right\rangle
\label{eq:interf2N}
\end{aligned}
\end{equation}
where
$
\Delta\Phi^{(m)}=\phi_{2N+1}-\phi_{2N-1}+2\omega_p\tau
+\Phi_-^{(m)}-\Phi_+^{(m)}$ is the single-shot interference phase and $\phi_{2N-1}$ and
$\phi_{2N+1}$ are the spectral phases of the two neighboring odd
harmonics, including
the deterministic HHG phase offset $\phi_0$ (see Supplementary Information). Equation \ref{eq:interf2N} encapsulates the core principle of the interferometric scheme for resolving phase correlations.  In the absence of phase correlations, random relative phases average out the interference over many realizations, completely washing out the oscillations.
\begin{figure}[h]
    \centering
\includegraphics[width=1\linewidth]{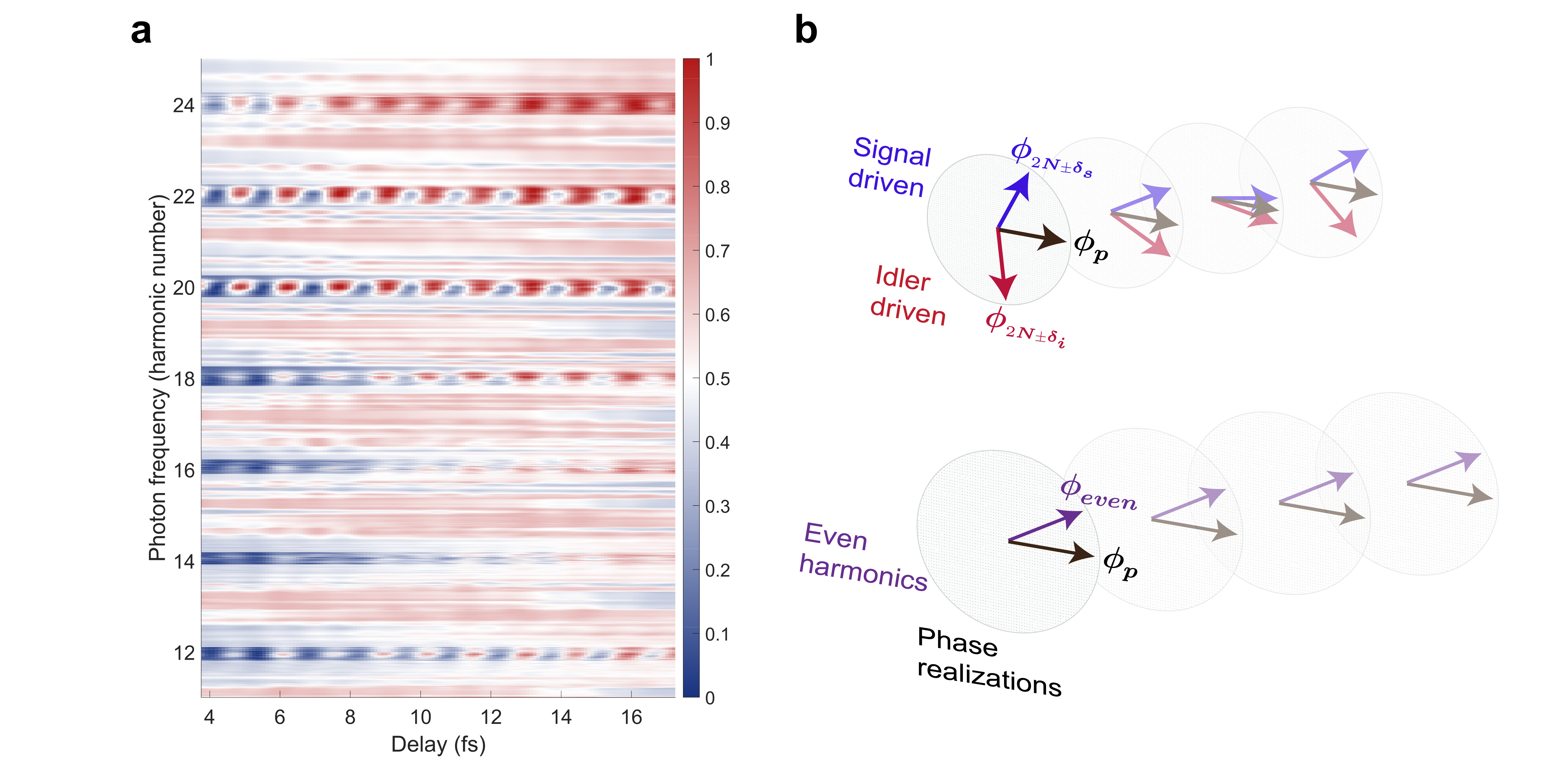}
\caption{
\textbf{Attosecond  correlation interferometry}
    {\bf (a)} Mean-value oscillations of the perturbed harmonic spectrum as a function of the pump$-$TMBSV delay, normalized per optical frequency.
    As can be clearly observed, no oscillations appear in the fractional harmonics, as their linear dependence on the TMBSV field, in the weak-perturbation regime, yields a single contribution rather than two interfering pathways. In contrast,  prominent oscillations at frequency $2\omega_p$ emerge across all even harmonics, revealing their interferometric nature. This measurement constitutes an attosecond complex-correlation interferometer between two quantum pathways, each composed of fractional harmonics, resolving both their amplitude and phase correlations. {\bf{(b)}} Resolving phase correlations via two-quantum-path interference. Even harmonics are generated through the interference of two quantum paths, each involving signal driven and idler driven fractional harmonics. The schematic illustrates the joint phase correlation between the two harmonic families. Red (blue) unit vectors represent the normalized phases of the idler driven (signal driven) harmonics for different experimental realizations, while the black vector denotes the pump-phase reference. Although the phases of the signal and idler driven harmonic groups fluctuate randomly from shot to shot, their sum remains locked to the pump phase. The two-path interferometer directly reveals this underlying phase correlation. }
\label{fig:phase_correlations_and_interference_pathways}
\end{figure}
We perform attosecond interferometry of  phase correlations by recording the mean harmonic response as a function of the pump–TMBSV delay. Figure~3a shows the DC-subtracted  mean HHG signal as a function of this delay. As can be clearly seen, the fractional harmonics exhibit a nearly flat response, remaining essentially constant with delay. Conversely, the even harmonics display pronounced oscillations with a frequency of $2\omega_p$. This behavior is consistent with the perturbative picture (see Supplementary Information): scanning the IR–TMBSV delay introduces a phase shift to each fractional harmonic without affecting its intensity. By contrast, the even-harmonic signal, arising from interference between two indistinguishable two-photon pathways, exhibits pronounced oscillations.

The delay-dependent even-harmonic oscillations provide a phase-sensitive measurement of the strong-field HHG pathways that generate the two families of fractional harmonics.  Although the fractional harmonics are individually thermal, as dictated by the statistical properties of the signal and idler perturbations, and exhibit random shot-to-shot phases, the interferometric scheme combines them into a coherent signal whose phase survives ensemble averaging (Figure ~\ref{fig:phase_correlations_and_interference_pathways}b).  This directly confirms the phase-sum relation predicted by equation~\eqref{eq:frac harmonics}, demonstrating the transfer of quantum origin phase correlations from the infrared to the XUV: $\phi_{2N \pm \delta_s} + \phi_{2N \pm \delta_i} \approx \mathrm{Const}$. Combined with the XUV amplitude-covariance and $g^{(2)}(0)$ measurements, these results establish the transfer of both amplitude and phase correlations into the XUV. Moreover, the pronounced oscillatory behavior of the even harmonics requires shot-to-shot photon-number and phase correspondence within the same sub-cycle ionization window, effectively implementing an intrinsic attosecond coincidence detector between the fractional harmonics.

Resolving the correlated spectral structure of the fractional harmonics naturally raises a central question: how are these correlations encoded in time? In particular, what is the temporal structure associated with the complex spectral correlations?
To address this question, we trace the fluctuating HHG signal into the time domain and reconstruct the temporal emission associated with the two groups of fractional harmonics, $2N \pm \delta_s$ and $2N \pm \delta_i$. The interferometric measurement, which provides correlated amplitude and phase information, serves as the basis for this reconstruction. For each family, spanning a broad spectral range, we perform a single-shot Fourier reconstruction, mapping spectral fluctuations into temporal fluctuations of the emitted attosecond pulse trains. Importantly, this procedure does not yield a deterministic field, but rather a statistical temporal object, whose structure reflects the underlying field statistics.

The time-domain reconstruction of the correlated attosecond pulse trains proceeds through the following steps. First, guided by their perturbative dependence on the signal and idler fields, the fractional harmonics are modeled as stochastic thermal fields. Second, our interferometric measurement (Figure~\ref{fig:phase_correlations_and_interference_pathways}) demonstrates that the fractional harmonics preserve the phase correlation:  $\phi_{2N\pm\delta_s}^{(m)} + \phi_{2N\pm\delta_i}^{(m)} \approx  \mathrm{Const}$. Finally, we estimate the mean spectral phase from the interferometric measurement at the even harmonics. The spectral phase of the high harmonics (and in particular the attochirp) is governed by the strong-field interaction and has been extensively studied \cite{ATTOCHIRPmairesse2003attosecond}. In our experiment, the attochirp is determined using the in-situ approach \cite{dudovich2006measuring,kim2013petahertz}.
As established in previous studies, perturbing the HHG process encodes the attochirp in the oscillation phase of the even harmonics. We therefore use the measured phase of these oscillations to reconstruct the spectral phase across the HHG spectrum. We note that our measurement is sensitive to spectral-phase fluctuations, which would otherwise wash out the even-harmonic oscillations. Combining this reconstruction with the stochastic spectral phases, phase correlations, and single-shot spectral amplitudes of the fractional harmonics yields the complex HHG spectrum and hence the attosecond field. Details of the reconstruction procedure are provided in the Supplementary Information.

Figure \ref{fig:time domain reconstruction} presents the temporal reconstruction of the signal and idler driven fractional-harmonic fields, both individually and jointly. When reconstructed separately, each field forms a stochastic attosecond pulse train confined within a femtosecond-scale envelope (Figure~\ref{fig:time domain reconstruction}a-d). However, because each field acquires a random thermal phase from shot to shot, the attosecond field structure averages out, and no stable sub-cycle oscillation is observed across shots. We next reconstruct the joint field by coherently combining the two fractional-harmonic groups (Figure~\ref{fig:time domain reconstruction}e,f). In striking contrast to the individual reconstructions, the joint field exhibits a pronounced attosecond-scale noise pattern, with $\sim$  30 attoseconds separating regions of maximal field fluctuations from nearly silent temporal windows, with a ratio of $\sim 6.8$ between the standard deviations of maximally noisy and maximally silent regions. This reconstruction establishes the mapping of femtosecond stochastic correlations in the TMBSV onto the attosecond regime, demonstrating that HHG transfers quantum-origin correlations across three orders of magnitude in timescale.

\begin{figure}[h]

    \centering
    \includegraphics[width=0.8\linewidth]{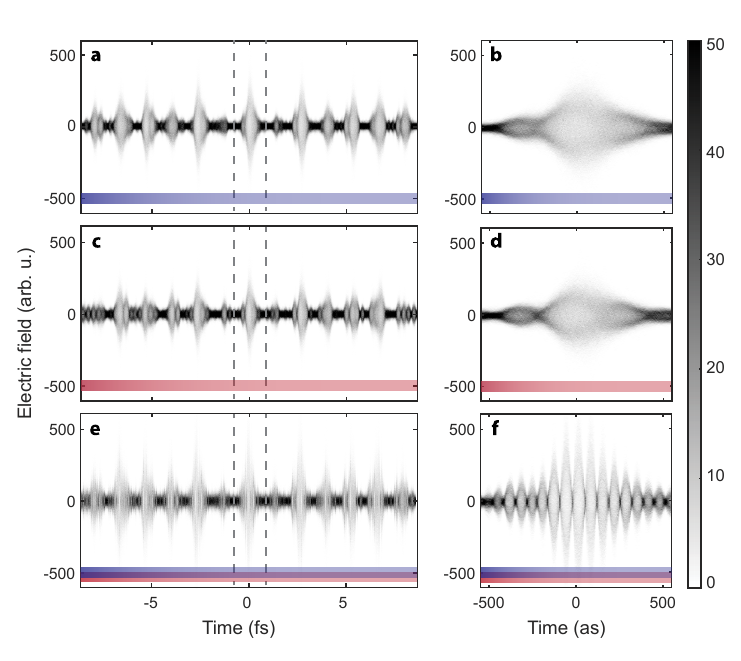}
\caption{\textbf{Time-domain reconstruction of correlated attosecond pulses.} Reconstructed temporal histograms obtained from the measured shot-to-shot fractional-harmonic families.  Time-domain reconstruction of signal driven fractional harmonics \textbf{(a,b)} and  idler driven fractional harmonics \textbf{(c,d)} across femtosecond and attosecond time scales, respectively. Each exhibits a femtosecond envelope, with blurred sub-cycle structure arising from stochastic phase fluctuations. \textbf{(e,f)}
Joint reconstruction of the total field of the two fractional-harmonic modes on femtosecond and attosecond timescales. A pronounced attosecond-scale structure is revealed, originating from both amplitude and phase correlations between the two XUV modes.
}
\label{fig:time domain reconstruction}
\end{figure}

In summary, in this study we demonstrate the generation, control, and measurement of quantum-origin correlations in XUV attosecond pulses. Using TMBSV to perturb HHG, we demonstrate that high-harmonic generation maps not only fluctuating spectral content from femtosecond to attosecond timescales, but also inter-mode correlation structures of quantum origin. We further demonstrate that even-harmonic generation acts as a sub-cycle coincidence measurement, enabling interferometric measurement of phase correlations with attosecond precision. Our study establishes a route towards quantum attosecond science. The ability to generate and probe quantum correlations in the XUV regime opens opportunities for quantum optics and quantum information at ultrafast timescales, extending concepts of non-classical light into the attosecond domain. Correlation-resolved XUV imaging may further provide new contrast mechanisms for ultrafast spectroscopy and microscopy. More broadly, our approach paves the way for the direct observation and control of correlated electron dynamics in matter, granting access to many-body interactions and electronic correlations on their natural attosecond timescale.

\section*{{\bf Acknowledgments}}
 N.D. is the incumbent of the Robin Chemers Neustein Professorial Chair. N.D. acknowledges the Minerva Foundation, the Israeli Science Foundation and the European Research Council for financial support. We thank Yossi Pilas for his technical support.

\paragraph*{{\bf Author contributions:}}
C.M. and N.D. supervised the study. A.S., K.D., C.M., N.D. and N.Y. conceived and designed the experiments. A.S., K.D., C.M. and N.Y. built the experimental setup. A.S., K.D., C.M., and N.Y. performed the experiments. A.S.,K.D. and C.M. analyzed the data. All authors interpreted the experimental and theoretical results, discussed the results and contributed to the final manuscript.
\paragraph*{{ \bf Competing interests:}}
There are no competing interests to declare.
\paragraph*{{ \bf Data and materials availability:}}
The data supporting the findings of this study are available upon reasonable request.

\clearpage

\bibliography{bibNow}
\bibliographystyle{unsrt}
\end{document}